\documentclass[a4paper]{jpconf}
\usepackage{graphicx}
\usepackage{amsmath}
\usepackage{bm}
\begin{document}
\title{Final-state interactions in deep-inelastic scattering from a tensor 
polarized deuteron target.}

\author{Wim Cosyn}

\address{Department of Physics and Astronomy, Ghent University, Proeftuinstraat 
86, 9000 Gent, Belgium}
\ead{wim.cosyn@ugent.be}

\author{Misak Sargsian}
\address{Department of Physics, Florida International University,
	Miami, Florida 33199, USA}

\begin{abstract}
Deep-inelastic scattering (DIS) from a tensor polarized deuteron is sensitive 
to possible non-nucleonic components of the deuteron wave function.  To 
accurately estimate the size of the nucleonic contribution, final-state 
interactions (FSIs) need to be accounted for in calculations.  We outline a 
model that, based on the diffractive nature of the effective hadron--nucleon 
interaction, uses the generalized eikonal approximation to model the FSIs in 
the resonance region, taking into account the proton-neutron component of the 
deuteron.  The calculation uses a factorized model with a basis of three 
resonances with mass $W<2$~GeV  as the relevant set of effective hadron states 
entering the final-state interaction amplitude for inclusive DIS.  We present 
results for the tensor asymmetry observable $A_{zz}$ for kinematics accessible 
in experiments at Jefferson Lab and Hermes.  For inclusive DIS, sizeable 
effects are found when including FSIs for Bjorken $x>0.2$, but the overall size 
of $A_{zz}$ remains small.  For tagged spectator DIS, FSIs effects are largest 
at spectator momenta around 300 MeV and for forward spectator angles.
\end{abstract}

\section{Introduction} \label{sec:intro}
Deep-inelastic scattering (DIS) from a polarized spin 1 target yields access to 
new structure functions not present for the spin $1/2$ case 
\cite{Hoodbhoy:1988am}.  The leading twist structure function $b_1$ is 
accessible in DIS from 
a tensor polarized spin 1 target with an unpolarized beam.  In the parton 
model, $b_1$ is sensitive only 
to the spin of the embedding spin 1 hadron.  For a spin 1 system of two 
non-interacting spin $1/2$ particles at rest, $b_1$ is trivially zero.  For the 
deuteron, due to the $D$-wave admixture in the wave function of a few percent, 
$b_1$ is still expected to be small $(b_1 \ll F_1^D)$ in the plane-wave 
approximation.  At low Bjorken $x<0.1$, shadowing effects are expected to yield 
significant contributions to $b_1$ 
\cite{Nikolaev:1996jy,Edelmann:1997qe,Bora:1997pi}. Measurement of 
a sizeable value for $b_1$ at moderate $x$ would provide insight into possible 
non-nucleonic components in the deuteron \cite{Miller:2013hla}.  The Hermes 
collaboration measured $b_1$ for $0.01<x<0.45$ at $0.5 < Q^2 < 5$ GeV$^2$, 
finding non-zero values and the measured tensor asymmetry $A_{zz}$ [see 
Eq.~(\ref{eq:cross}) below] exhibited a sign change around $x \approx 0.3$ 
\cite{Airapetian:2005cb}.  An experiment at Jefferson Lab (JLab) will provide a 
new measurement for $b_1$ at $0.16<x<0.5$ and $1<Q^2<5$ GeV$^2$ 
\cite{Slifer:2013vma}.

So far, no study has been made of possible final-state interaction (FSI) 
effects at moderate $x$ to the nucleonic contribution of $b_1$.  It is 
important to quantify this contribution as a sizeable FSI effect could 
dominate possible non-nucleonic contributions.  Here, we present work in a 
model that has been used to account for FSIs in DIS from the deuteron for 
inclusive \cite{Cosyn:2013uoa} and tagged spectator measurements 
\cite{Cosyn:2010ux,Cosyn:2011jc}.  The major difficulty in describing FSI in 
these reactions is that one lacks detailed
information about the composition and space-time evolution of the hadronic
system produced in the deep inelastic scattering and how this changes as a
function of Bjorken $x$ and $Q^2$.  

\section{Formalism} \label{sec:model}
The model introduced in Ref. \cite{Cosyn:2010ux} accounted
for FSI effects in DIS from the deuteron based on general properties of 
high-energy diffractive scattering. The underlying assumption was that due to 
the restricted
phase space (finite values of $W$ and $Q^2$), the minimal Fock state
component of the wave function can be used to describe DIS from the
bound nucleon.  In this case the scattered state consists of three
outgoing valence quarks whose rescattering from the spectator nucleon
is parametrized in the form of a $Q^2$- and $W$-dependent diffractive
amplitude.

The model uses the virtual nucleon 
approximation (VNA) \cite{Melnitchouk:1996vp,Sargsian:2005rm,Sargsian:2001gu}
to describe DIS from the deuteron.  In the VNA, only the 
proton-neutron component of the deuteron system is considered.  It is also 
assumed that the negative energy projection of the virtual nucleon propagator 
gives negligible contributions to the scattering amplitude and the 
contribution of meson exchange currents is neglected.
In the VNA, the nuclear  wave  function  $\Psi_D(p)$ is  normalized to account 
for baryon
number 
conservation \cite{Frankfurt:1981mk}:
\begin{equation}
\int \alpha |\Psi_D(p)|^2 d^3p  = 1,
\label{norm}
\end{equation}
where $\alpha= 2-\frac{2(E_s-p_{s,z})}{M_D}$ is the light cone momentum
fraction of the deuteron 
carried by the bound nucleon, with $E_s (p_s)$ the on-shell energy (momentum) 
of the spectator nucleon.
Because of the virtuality of the interacting nucleon it is impossible 
to satisfy the momentum  sum rule at the same time, which can be qualitatively 
interpreted as part  of the deuteron momentum
fraction being distributed to non-nucleonic degrees of freedom.

The final-state interaction contribution is encoded in a Feynman diagram where 
after the interaction of the virtual photon with the bound nucleon, the 
produced hadronic mass $X$ interacts with the spectator nucleon through an 
effective rescattering amplitude.  At the energies for the experiments 
considered here, we can assume that the rescattering will be highly diffractive 
and will occur over small angles and the generalized eikonal approximation 
(GEA) \cite{Sargsian:2001ax} can be applied.  In the GEA an eikonal form is 
adopted for the effective rescattering amplitude:
\begin{equation} \label{eq:fsiamp}
 f_{X^\prime N, X N} = \sigma(Q^2,x)(i + \epsilon(Q^2,
x))e^{\frac{B(Q^2,x)}{2} t},
\end{equation}
where  $\sigma$, $\epsilon$ and $B$ are the effective total cross section, real
part and the slope factor 
of the diffractive $X^\prime N\rightarrow X N$ scattering amplitude.  The $x$ 
and $Q^2$ dependence of the scattering parameters $\sigma$ and $B$ were 
determined by fitting our model calculations to data for tagged spectator 
deuteron DIS taken in the Deeps experiment at Jefferson Lab 
\cite{Cosyn:2010ux,Klimenko:2005zz}.
In the derivation of the FSI amplitude, a factorized approach is used, whereby
the interaction of the virtual photon with the off-shell nucleon (encoded in 
the nucleon structure functions) is taken out
of the integration over the intermediate spectator momentum.  Finally, the four 
structure functions for tagged spectator deuteron DIS ($F_T,F_L,F_{LT},F_{TT}$) 
can be written as the product of a distorted deuteron momentum distribution 
containing the amplitude of Eq.~(\ref{eq:fsiamp}), the nucleon structure 
function $F_2^N$ and kinematical factors (for detailed expressions see Ref. 
\cite{Cosyn:2010ux}).  

By applying the optical theorem to the forward virtual Compton scattering 
amplitude, the model was extended to inclusive deuteron DIS in 
Ref.~\cite{Cosyn:2013uoa} :
\begin{equation} 
 W^{\mu\nu}_D
= \frac{1}{2 \pi M_D} \frac{1}{3}
  \sum_{s_D} {\Im}m\, {\cal A}^{\mu\nu}_{\gamma^* D}(t=0)\,,
\label{eq:optical}
\end{equation}
where $W^{\mu\nu}_D$ is the hadronic tensor of the inclusive DIS process and 
${\cal A}^{\mu\nu}_{\gamma^* D}$ the virtual Compton scattering amplitude.  The 
advantage
of such an approach is that the amplitudes accounting for the FSI
effects will self-consistently satisfy the unitarity conditions for
inelastic rescattering.
The 
amplitude ${\cal A}^{\mu\nu}_{\gamma^* D}$ contains contributions from the 
plane-wave diagram and the rescattering diagram, where the produced hadronic 
mass $X$ interacts with the spectator nucleon.  Again, a factorized 
approximation is applied and for the intermediate hadronic states three 
effective resonances (at invariant mass $W=1.232, 1.5,$ and 1.75 GeV) were 
taken into account in addition to a broad contribution from the DIS continuum.  
We showed that in our formalism, FSI effects naturally disappear as $Q^2$ 
increases due to phase-space constraints and found sizeable FSI effects for 
$x>0.6$ and $Q^2<10 \text{GeV}^2$.

It is now straightforward to extend the model to the situation of a polarized 
deuteron target.  It suffices to replace the average over the deuteron 
polarization by a trace with the appropriate spin 1 density matrix.  
Consequently, Eq.~(\ref{eq:optical}) can be written as
\begin{equation}
 W^{\mu\nu}_D
= \frac{1}{2 \pi M_D} \sum_{\lambda \lambda'} {\Im}m\,  \rho^D_{\lambda' 
\lambda}
\left[ {\cal A}^{\mu\nu}_{\gamma^* D}(t=0) \right]_{\lambda \lambda'}\,, 
\end{equation}
where $\rho^D$ represents the spin 1 density matrix for the deuteron.

\section{Deuteron density matrix}\label{sec:densitymatrix}
The density matrix for a spin 1 particle in a rest frame for the particle can 
be written in a multipole expansion \cite{Leader:2001gr}
\begin{equation}\label{eq:mult}
 \rho^D_{\lambda \lambda'}=\frac{1}{3}\sum_{L=0}^2\sum_{M=-L}^L(2L+1)t_{LM}^* 
\left(\hat{T}^L_M\right)_{\lambda \lambda'}\,,
\end{equation}
where $\lambda$ denotes the spin projection of the deuteron, $t_{LM}$ are 
the multipole parameters, and 
the spherical tensor operators read $\left(\hat{T}^L_M\right)_{\lambda 
\lambda'} = \langle s\lambda|s\lambda';LM \rangle$.  When the ensemble is made 
up of particles quantized along the $z$-axis, the density matrix takes on a 
simple form
\begin{equation}\label{eq:dens}
 \rho^D = 
\frac{1}{3} 
\text{diag}(1+\frac{3}{2}P_z+2P_{zz},1-P_{zz},1-\frac{3}{2}P_z+2P_{zz })\,,
\end{equation}
where $P_z=\sqrt{2}t_{10}=p^+-p^-$ and $P_{zz}=\sqrt{10}t_{20}=p^++p^--2p^0$, 
with $p^\lambda$ the probability of finding the deuteron with a spin projection 
$\lambda$ along the $x$-axis.

In theoretical calculations for deuteron DIS, the hadronic tensor is 
usually evaluated in the hadron plane (determined by the virtual photon and 
spectator nucleon), with the $z$-axis along the virtual photon momentum.  For a 
fixed target experiment, however, the deuteron density matrix is usually 
determined in a deuteron rest frame with the $z$-axis along the incoming beam.  
Consequently, care has to be taken to transform the density matrix to the hadron 
plane in order to do meaningful comparisons with experimental data.  This 
involves a rotation over the angle between the incoming beam and virtual photon 
$\theta_{eq}$, and over the angle $\phi$ between the electron and hadron planes. 
 Due to the simple transformation properties under rotations of the multipole 
parameters in Eq.~(\ref{eq:mult}), this transformation can be readily 
computed \cite{Leader:2001gr}.  For an ensemble with only the multipole 
parameter $t_{20}$ nonzero in the rest frame with the $z$-axis along the 
incoming beam, one obtains for the multipole parameters $\hat{t}_{LM}$ in the 
hadron plane \cite{Jeschonnek:2009tq}:
\begin{align}\label{eq:pol}
 \hat{t}_{20} &= \frac{1}{4}(1+3\cos 2\theta_{eq}) \,t_{20} \nonumber \,, \\
 \hat{t}_{21} &= -\sqrt{\frac{3}{8}} \sin 2\theta_{eq} e^{i\phi} \,t_{20} 
\nonumber \,, \\
 \hat{t}_{22} &=  \sqrt{\frac{3}{32}}(1-\cos 2\theta_{eq})e^{i2\phi} \,t_{20} 
 \,.
\end{align}
For the deuteron DIS formalism under consideration here, this means that even 
after integrating over the angle $\phi$ (which is done in the inclusive DIS 
formalism), the cross section will receive contributions from the $F_{LT}$ and 
$F_{TT}$ response functions (that come with a $\cos\phi$ and $\cos2\phi$ 
dependence) through respectively the  $\hat{t}_{21}$ and $\hat{t}_{22}$ parts 
of the density matrix.

\section{Results} \label{sec:results}
The cross section for DIS from a polarized deuteron target without any beam 
polarization can be written as
\begin{equation}\label{eq:cross}
 d\sigma = d\sigma_u ( 1+\frac{1}{2}P_{zz}A_{zz}) \,,
\end{equation}
where $\sigma_u$ is the cross section for an unpolarized target, $P_{zz}$ was 
defined in Eq.~(\ref{eq:dens}), and $A_{zz}$ is the 
tensor asymmetry of the deuteron cross section which can be related to the 
structure function $b_1$ \cite{Airapetian:2005cb}.  For the calculations of 
$A_{zz}$ for inclusive DIS from the deuteron, we include resonances at invariant 
mass $W=1.232, 1.5,$ and 1.75 GeV as the set of effective hadron states entering
the final-state interaction amplitude. The values of the scattering 
parameters in Eq.~(\ref{eq:fsiamp}) are taken from our fits to the JLab Deeps 
data.  We used the SLAC nucleon structure function parametrization 
\cite{PhysRevD.20.1471} and for the non-relativistic deuteron wave function,
we use the parametrization based on the Paris $NN$ potential
\cite{Lacombe:1980dr}.  

\begin{figure}[ht]
\includegraphics[width=18pc]{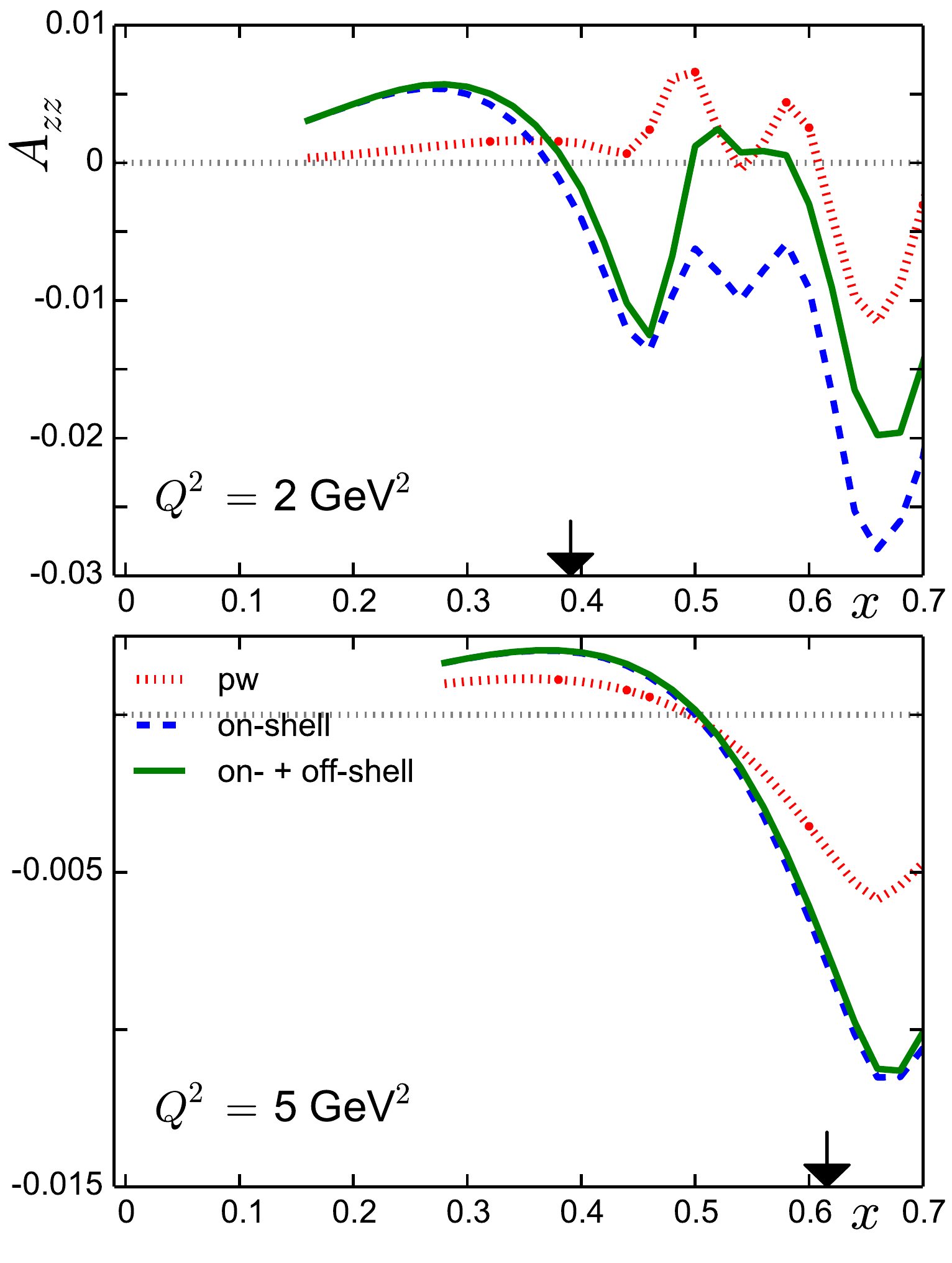}\hspace{2pc}%
\includegraphics[width=18pc]{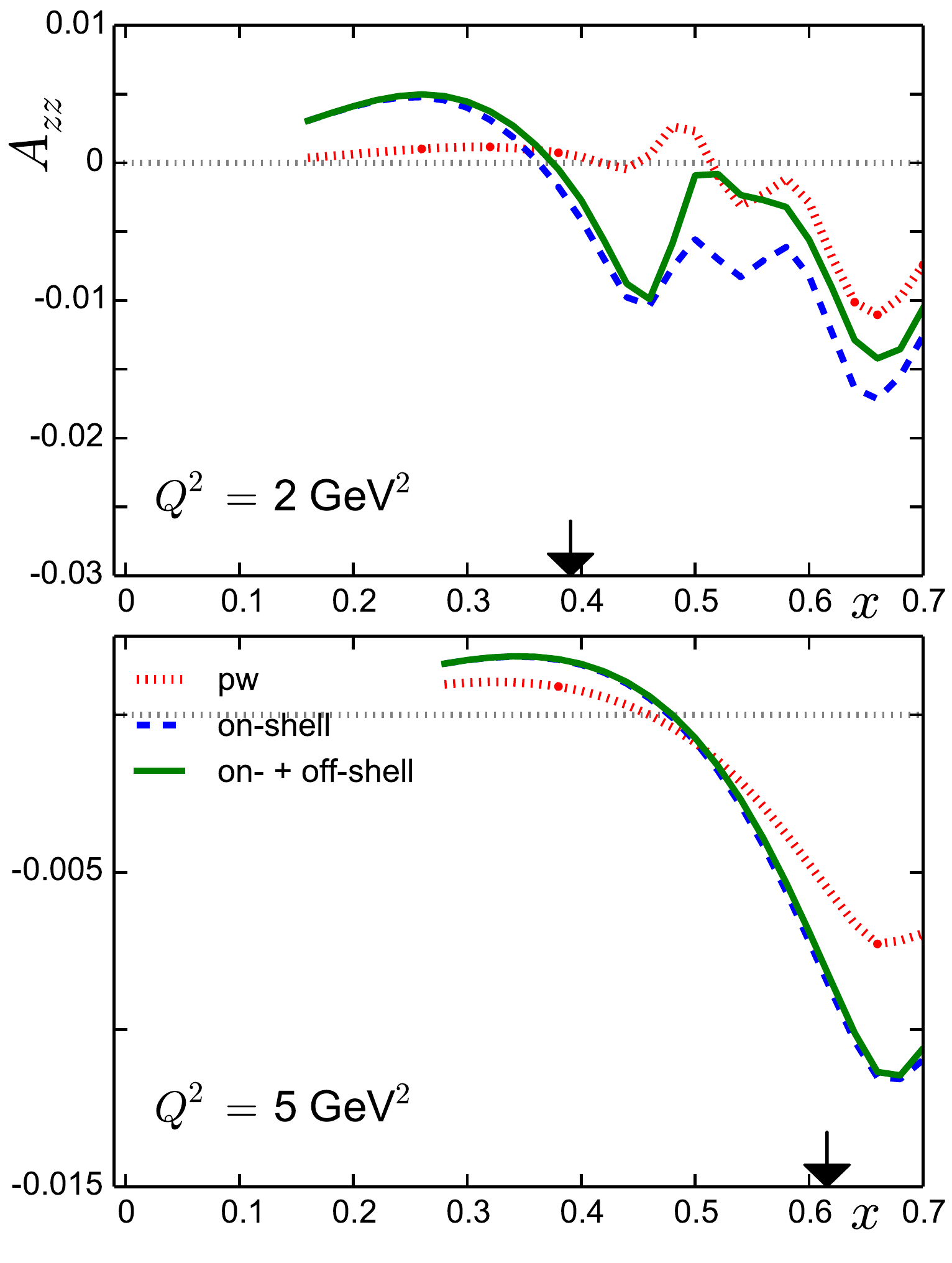}\hspace{2pc}%
\caption{\label{fig:inc} Tensor asymmetry $A_{zz}$ for kinematics accessible in 
the upcoming JLab experiment \cite{Slifer:2013vma}  (incoming beam of 11 GeV). 
The red dotted curve shows the plane-wave calculations, the blue dashed 
(green full) curve includes the contribution of the on-shell (and off-shell) 
FSI 
contribution.  For details on the FSI on- and off-shell calculations, we refer 
to Ref.~\cite{Cosyn:2013uoa}.  Left panels have the deuteron polarized along the 
virtual photon direction, right panels along the incoming beam, with the 
transformation according to Eq.~(\ref{eq:pol}). The arrow along the $x$-axis 
indicates the boundary at
	$W=2$~GeV between the resonance and DIS regions for free
	nucleon kinematics.}
\end{figure}

In Fig.~\ref{fig:inc}, we 
present calculations of $A_{zz}$ for inclusive DIS from the deuteron for 
kinematics accessible in the upcoming JLab experiment 
\cite{Slifer:2013vma}.  For the plane-wave calculations, we observe that 
$A_{zz}$ is almost zero for Bjorken $x$ in the DIS region, while it reaches 
values of the order of $\sim \pm0.01$ in the resonance region at $Q^2=2~ 
\text{GeV}^2$ and becomes smaller with increasing $Q^2$.  Adding the FSI 
contributions has a significant effect on the size of $A_{zz}$ over the whole 
$x$ range.  At $Q^2=2~\text{GeV}^2$ the off-shell contribution of the FSI 
diagram also has a sizeable contribution while at $Q^2=5~\text{GeV}^2$ it is 
almost negligible.  It is worth remarking that even though the effective 
hadronic states taken into account in the FSI diagram all lie in the resonance 
region, they also contribute significantly to $A_{zz}$ in the DIS region.  This 
can be understood in the following manner.  In the formalism $A_{zz}$ is only 
nonzero because of the $D$-wave component of the deuteron wave function, both 
in the plane-wave and FSI contribution to the cross section.  As the dominant 
contribution for the $D$-wave occurs at momenta above 250 MeV, $A_{zz}$ is 
still sensitive to the resonance region FSI contributions in the DIS $x$-region 
for free nucleon kinematics due to the Fermi motion of the struck nucleon.  
Comparing the effect of tensor polarization along the virtual photon with 
polarization along the incoming beam, we find limited differences for 
$Q^2=5~\text{GeV}^2$. For $Q^2=2~\text{GeV}^2$ the size of $A_{zz}$ 
decreases, especially at the largest $x$ values, where the angle $\theta_{eq}$ 
is largest.

\begin{figure}[ht]
\includegraphics[width=18pc]{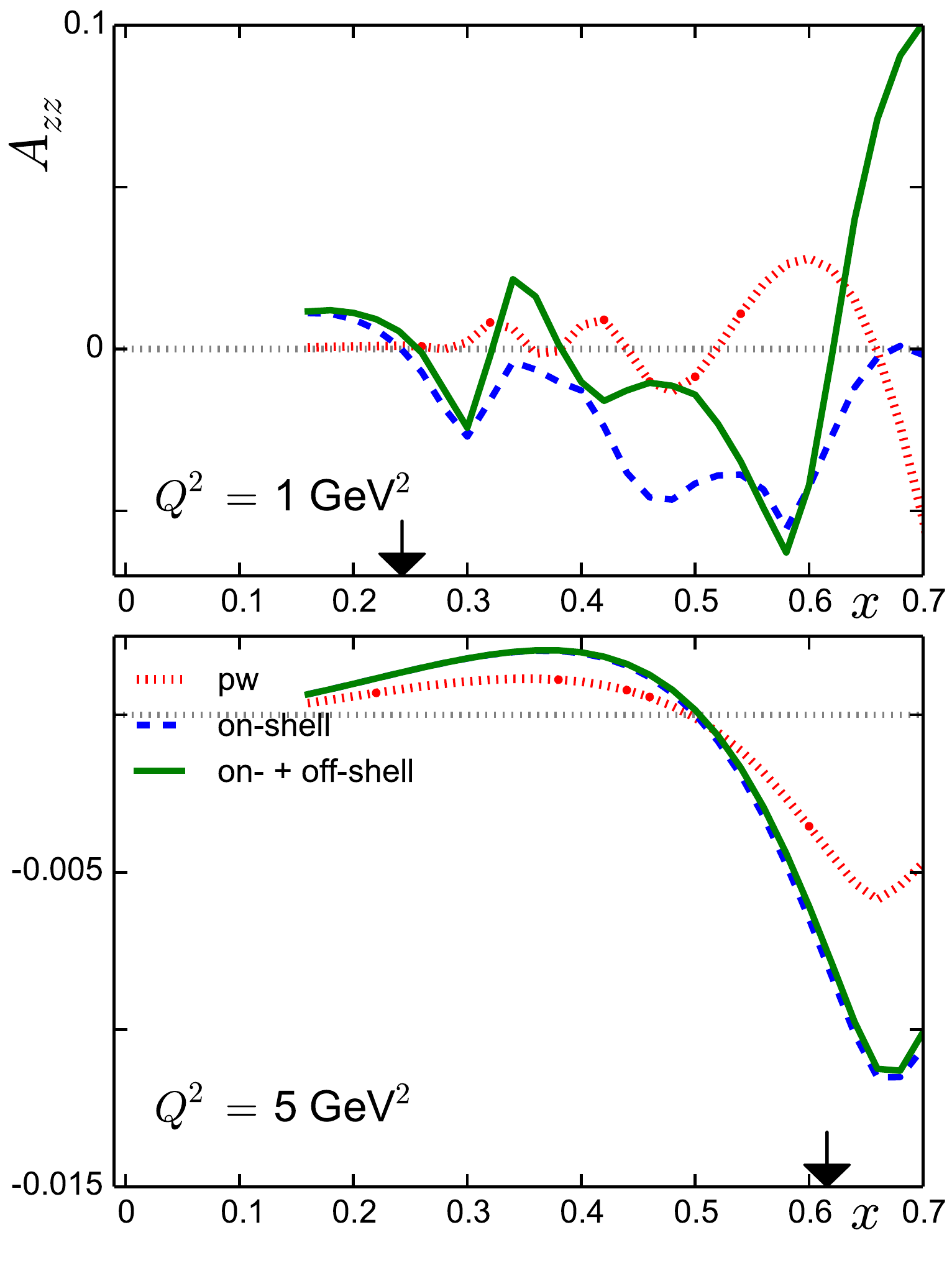}\hspace{2pc}%
\includegraphics[width=18pc]{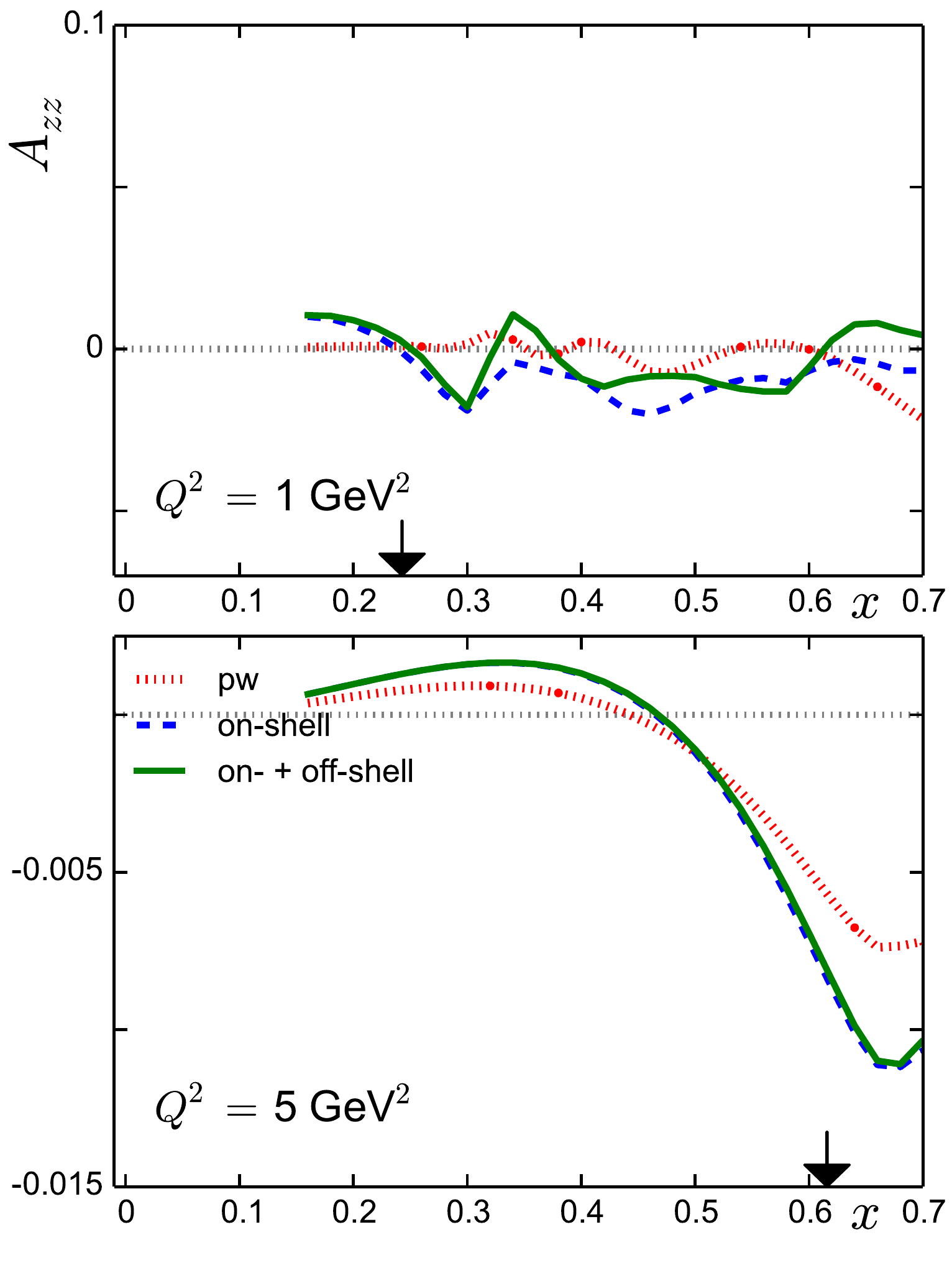}\hspace{2pc}%
\caption{\label{fig:inc2} Tensor asymmetry $A_{zz}$ for kinematics covered in 
the Hermes experiment \cite{Airapetian:2005cb}  (incoming beam of 27.6 GeV). 
Curves and panels as in Fig.~\ref{fig:inc2}}
\end{figure}

In Fig.~\ref{fig:inc2}, we show similar $A_{zz}$ calculations as in 
Fig.~\ref{fig:inc} but for kinematics covered in the Hermes experiment 
\cite{Airapetian:2005cb}.  Similar observations as 
in Fig.~\ref{fig:inc} apply. The difference between the left and right panel 
for $Q^2=1~\text{GeV}^2$ is more pronounced here because of larger 
$\theta_{eq}$ values, with $A_{zz}$ values including FSIs becoming significantly 
smaller for the case of polarization along the incoming beam.  Comparing our 
calculation to the Hermes data point of $A_{zz}=0.157\pm 0.69$ at $x=0.45$, 
$Q^2\approx 5 \text{GeV}^2$, we find a value of $A_{zz} \approx 0.0015$, about 
two orders of magnitude smaller.

\begin{figure}[ht]
\begin{center}
\includegraphics[width=30pc]{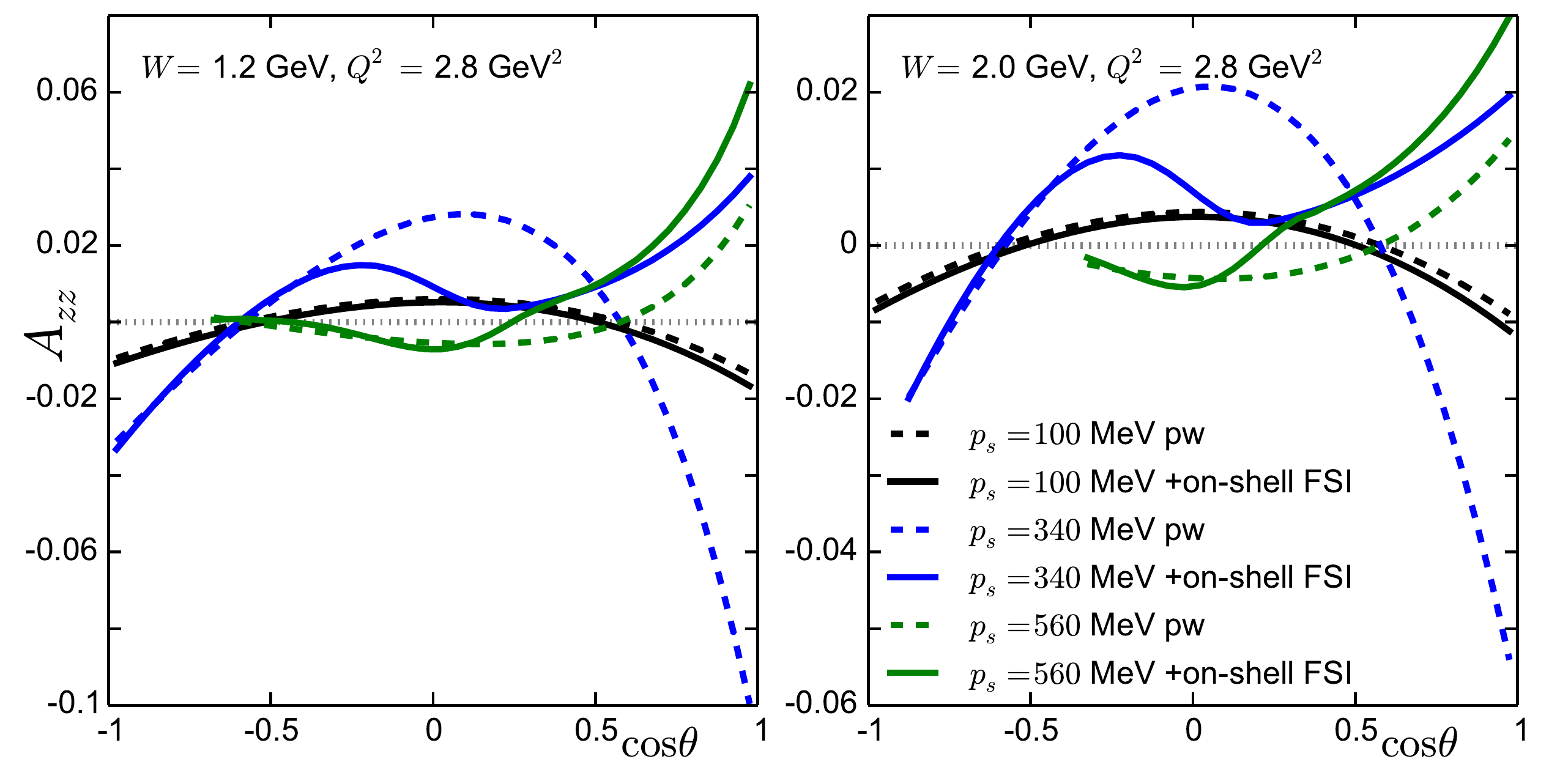}\hspace{2pc}%
\end{center}
\caption{\label{fig:sidis}Tensor asymmetry for tagged 
spectator DIS $D(e,e'p_s)X$ at invariant mass $W$ of $X$ 1.2 (left panel) and 
2.0 GeV (right panel) for three different spectator momenta.  Full (dashed) 
curves represent a plane-wave (including FSIs) calculation.  For details on the 
FSI calculations, see Ref.~\cite{Cosyn:2010ux}.}
\end{figure}

Finally, in Fig.~\ref{fig:sidis}, we show calculations for tagged spectator DIS 
[$D(e,e'p_s)X$] from the deuteron at different spectator momenta and two 
values of the invariant mass $W$ of the produced $X$, for a deuteron polarized 
along the virtual photon momentum.  One observes that the plane-wave 
calculations produce sizeable values for $A_{zz}$ in the case of intermediate 
spectator momenta of 340 MeV (where the deuteron $D$-wave contribution is 
largest), especially for forward spectator angles.  Including FSIs has little 
effect at the low spectator momentum of 100 MeV, but causes significant changes 
for higher spectator momenta in the forward region, where we know from our 
previous study \cite{Cosyn:2010ux} that FSI effects are largest in the 
unpolarized cross section.  For $p_s=340$ MeV and $\cos \theta_s>0.5$ including 
the FSI contribution even reverses the sign of $A_{zz}$.

\section{Conclusion} \label{sec:conclusion}
We presented calculations for the deuteron tensor asymmetry $A_{zz}$ in a model 
based on the virtual nucleon and generalized eikonal approximations that allows 
for the description of final-state interaction effects for deuteron DIS 
reactions from intermediate hadronic states in the resonance region.  For 
inclusive DIS, we included resonances at invariant mass $W=1.232, 1.5,$ and 1.75 
GeV as the set of effective hadron states entering
the final-state interaction amplitude. Our calculations for kinematics 
accessible at JLab12 and Hermes showed sizeable FSI 
effects both in the resonance and DIS region for free-nucleon kinematics, 
though the overall size of the nucleonic contribution to $A_{zz}$ remains 
small. The sensitivity of $A_{zz}$ in the DIS region to FSI contributions from 
the resonance region can be understood from the dependence of the $A_{zz}$ 
observable on the $D$-wave part of the deuteron wave function.  We showed 
that deuteron polarization along the incoming beam reduces the size of 
$A_{zz}$ at the low $Q^2$ values compared to polarization along the virtual 
photon direction.  For the tagged spectator DIS process FSI effects proved to 
be largest at spectator momenta around 300 MeV and for spectator angles $\cos 
\theta_s >0.5$.

\ack
The computational resources (Stevin Supercomputer Infrastructure) and
services used in this work were provided by Ghent University, the
Hercules Foundation and the Flemish Government – department EWI.  This
work is supported by the Research Foundation Flanders as well as by the  
U.S. Department of Energy Grant 
under Contract DE-FG02-01ER41172.

\section*{References}
\bibliography{../bibtexall.bib}

\end{document}